\documentclass[twocolumn,prb,showpacs,preprintnumbers,amsmath,amssymb,floatfix,superscriptaddress]{revtex4-1}
\usepackage{graphicx}
\usepackage{amsmath}
\usepackage{amssymb}
\usepackage[normalem]{ulem}
\usepackage{color}

\begin{document}

\title{Dimensional effects on the tunneling conductivity of gold-implanted nanocomposite films}

\author{C. Grimaldi}\affiliation{Laboratory of Physics of Complex Matter, Ecole Polytechnique F\'ed\'erale de
Lausanne, Station 3, CH-1015 Lausanne, Switzerland}
\author{M. Cattani}\affiliation{Institute of Physics, University of S\~ao Paulo, C. P. 66318, CEP 05314-970, S\~ao Paulo, S. P., Brazil}
\author{M. C. Salvadori}\affiliation{Institute of Physics, University of S\~ao Paulo, C. P. 66318, CEP 05314-970, S\~ao Paulo, S. P., Brazil}



\begin{abstract}
We study the dependence of the electrical conductivity on the gold concentration of Au-implanted 
polymethylmethacrylate (PMMA) and alumina nanocomposite thin films. 
For Au contents larger than a critical concentration, the conductivity of Au-PMMA and Au-alumina 
is well described by percolation in two dimensions, indicating that the critical correlation length for percolation is larger than the thickness of the films. Below the critical loading, the conductivity is dominated by
tunneling processes between isolated Au particles dispersed in PMMA or alumina continuous matrices.
Using an effective medium analysis of the tunneling conductivity, we show that Au-PMMA behaves as a 
tunneling system in two dimensions, as the film thickness is comparable to 
the mean Au particle size. On the contrary, the conductivity of Au-alumina films is best described by tunneling 
in three dimensions, although the film thickness is only a few times larger than the particle size. We interpret 
the enhancement of the effective dimensionality of Au-alumina films in the tunneling regime as due to the 
larger film thickness as compared to the mean interparticle distances. 
\end{abstract}
\maketitle

\section{Introduction}
\label{intro}

Nanocomposite films made of nano-sized metallic particles embedded in an insulating matrix display peculiar 
electron transport regimes depending on the relative amount of the metallic phase compared to the insulating one. 
Above a critical concentration   $x_c$ of the metallic phase, macroscopic clusters of connected metallic particles span the entire 
sample, giving rise to an electrical conductivity $\sigma$ characterized in the vicinity of $x_c$ by a power-law behavior of the form:
\begin{equation}
\label{power}
\sigma\propto (x-x_c)^t,
\end{equation}where $t$  is the transport exponent which the percolation theory predicts to take the values $t\simeq 1.3$  and $t\simeq 2$ for strictly 
two-dimensional (2d) and three-dimensional (3d) systems, respectively.\cite{Stauffer1994,Sahimi2003} 
Equation \eqref{power} describes well the conductivity behavior in the $x>x_c$  region of several nanogranular metal films grown 
by different methods,\cite{McAlister1985, Toker2003,Salvadori2008,Wei2013} with observed values of the exponent that are consistent 
with the effective dimensionality of the films. Below $x_c$  the metallic phase is broken up into disconnected metallic regions, so that 
for concentrations sufficiently smaller than $x_c$  electrons have to tunnel across the insulating barrier separating the conducting fillers. 
In this regime, the conductivity of nanocomposites with embedded, isolated metallic particles follows 
approximately:\cite{Ambegaokar1971,Hunt2005,Ambrosetti2010a,Ambrosetti2010b}
\begin{equation}
\label{tun1}
\sigma\propto e^{-\frac{2\delta(x)}{\xi}},
\end{equation} 
where $\xi$ is the tunneling decay length which, depending on the nature of the composite constituents, ranges from a fraction to a few 
nanometers, and $\delta(x)$  is a typical distance between the surfaces of two conducting particles. In the dilute limit $x\rightarrow 0$, 
dimensional considerations\cite{Note1} imply that $\delta(x)\propto D/x^{1/d}$  for particles of mean size $D$ randomly dispersed 
in a $d$-dimensional volume, as also inferred from transport measurements on 2d and 3d nanogranular metal composites.\cite{Wei2013,Fostner2014}

Recently, a survey of the conductivity data of several thick nanogranular composite films has evidenced that the percolation behavior 
of Eq.~\eqref{power} evolves into the tunneling one of Eq.~\eqref{tun1} as the result of the competition between percolation and 
tunneling transport mechanisms as  $x$ decreases.\cite{Grimaldi2014}

In this article we study the conductivity behavior as a function of gold content in Au-implanted nanocomposite films above and below  $x_c$. 
We consider two distinct series of systems: Au-implanted alumina and Au-implanted polymethylmethacrylate (PMMA) films. Au-alumina 
films have thicknesses  $h$ a few times larger than the mean Au particle size $D$, while the conducting layer of Au-PMMA is basically 
formed by a single layer of gold particles ($h\approx D$), which makes this system strictly two dimensional. 
We show that for $x>x_c$ the conductivity of both systems follow Eq.~\eqref{power} with transport exponents very close to the 2d value $t\simeq 1.3$, 
in accord with the prediction of the percolation theory of nanocomposites with nanometric values of $h$. To study the tunneling behavior 
at Au concentrations below $x_c$, we analyze the conductivities of Au-PMMA and Au-alumina using an effective medium theory formulated for 
film thicknesses ranging from $h=D$  to $h\gg D$, as to describe the evolution from 2d to 3d of the conductivity. We find that, in contrast to the 
2d nature of the Au-PMMA films, the Au-alumina films are large enough to sustain tunneling conductivity with 3d character. The effective 
dimensionality of our ion-implanted Au-alumina nanocomposites thus increases from 2d to 3d as the system crosses over from the 
percolation to the tunneling regimes.  This conclusion is supported by the value of $\xi$  extracted from the conductivity data, which coincides 
with that observed\cite{Grimaldi2014} in thick (i.e., 3d) co-sputtered Au-alumina nanocomposite films.\cite{Abeles1975}

\section{Material and methods}
\label{material}

The experimental data used in the present work are basically the same ones presented in Refs.~\onlinecite{Salvadori2008,Salvadori2013,Teixeira2009}, 
but with a completely different approach and exploring a different range of the data. The mentioned works addressed specifically 
implantation doses above the percolation thresholds, while the present work focuses in doses below the 
percolation thresholds, where the particles are isolated from each other and the tunneling effect is dominant.

A summary of the experimental procedure will be done here and details can be obtained in the mentioned references. 

PMMA was deposited on glass microscope slides using a spin coater, generating a film thickness about $50$ nm. Electrical contacts 
were formed at both ends of the substrates by plasma deposition of thick ($200$ nm) platinum films. Very low energy ($49$ eV) was 
used for ion implantation using a streaming (unidirectionally drifting) charge-neutral plasma formed by a vacuum arc plasma gun. 

Gold ion implantation in alumina 
using $40$ keV was proceeded in an implanter.\cite{Salvadori2012} Electrical contacts were 
formed at both ends of the alumina sample using silver paint. 

For both Au-PMMA and Au-alumina systems, in situ resistance measurements were performed as the ion implantation 
proceeded:\cite{Salvadori2008,Salvadori2013} after a known dose of Au ions is implanted, the implantation process is temporary halted 
and the resistance is measured. This process is repeated to determine the sample conductivity as a function of  ion implantation dose.

In the present work, computer simulation using the TRIDYN computer code\cite{Moeller1984,Moeller1988} was used to estimate 
the depth profiles of the ion implanted gold in the alumina substrate. TRIDYN is a Monte Carlo simulation program based on 
the TRIM (Transport and Range of Ions in Matter) code.\cite{Ziegler1985} The program takes into account compositional changes 
in the substrate due to: previously implanted dopant atoms, and sputtering of the substrate surface.
Note that, the parameters $t$ (transport exponent) and $x_c$  (critical concentration of the metallic phase) were recalculated in 
the present work, using a different approach, presenting some deviation from those obtained in the previous works.\cite{Salvadori2008,Salvadori2013}

\section{Results}
\label{results}

\begin{figure}[t]
\begin{center}
\includegraphics[scale=0.42,clip=true]{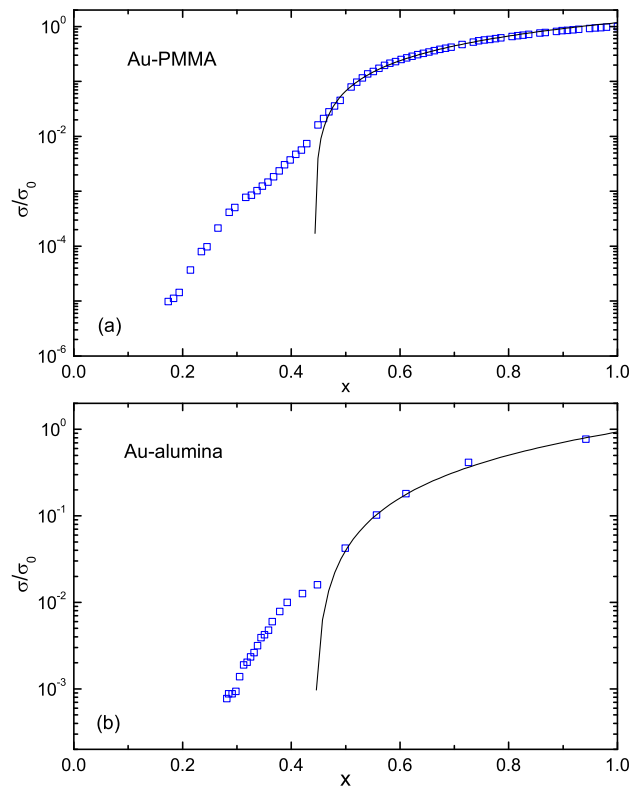}
\caption{(Color online) Normalized conductivity (open squares) of (a) Au-PMMA and (b) Au-alumina nanocomposite films 
as a function of the normalized ion dose $x$. Solid lines are fits with Eq.~\eqref{power}.  }\label{fig1}
\end{center}
\end{figure}

Results of our conductivity measurements on Au-PMMA and Au-alumina nanocomposites are shown in Fig.~\ref{fig1} (open squares), where 
the conductivity ratio $\sigma/\sigma_0$ is plotted as a function of the normalized Au concentration $x=\varphi/\varphi_0$. 
The saturation dose $\varphi_0$ is defined as the implantation dose $\varphi$ above which a continuous metal film starts to 
be deposited on the insulator surface, while the saturation conductivity $\sigma_0$ is the conductivity measured at 
$\varphi=\varphi_0$ (i.e., at $x=1$). We have determined $\varphi_0=2\times 10^{16}$ atoms cm$^{-2}$ and $\sigma_0=2\times 10 ^6$ S/m 
for Au-PMMA,\cite{Salvadori2008} and $\varphi_0=9.5\times 10^{16}$ atoms cm$^{-2}$ and $\sigma_0=14$ S/m for Au-alumina.\cite{Salvadori2013} 
As is apparent from the semi-log plot of Fig.~\ref{fig1}, the $x$-dependence of $\sigma/\sigma_0$ for both Au-PMMA and Au-alumina films 
presents a double hump shape, which is a feature commonly observed also in several other nanogranular metals composites.\cite{Grimaldi2014}
The hump at values of $x$ larger than $0.4-0.5$ stems from the presence of clusters of coalesced Au particles that extent 
across the entire composite layer. In this region, $\sigma/\sigma_0$ is reasonably well fitted by Eq.~\eqref{power} (shown 
by solid lines in Fig.~\ref{fig1}) with $t=1.26\pm 0.03$ and $x_c=0.443\pm 0.002$ for Au-PMMA, and $t=1.4\pm 0.2$ and $x_c=0.44\pm 0.02$ 
for Au-alumina composite films. 

\begin{figure}[t]
\begin{center}
\includegraphics[scale=0.42,clip=true]{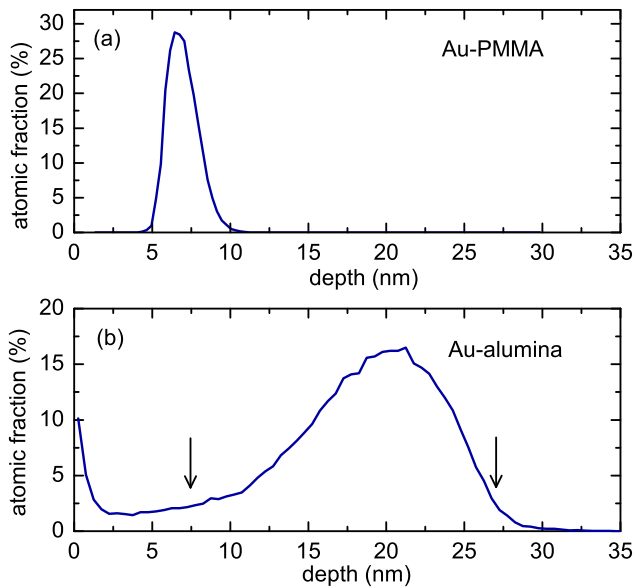}
\caption{(Color online) TRIDYN simulations of the depth profiles for gold implanted into (a) PMMA with ion energy $49$ eV and 
dose $\varphi=0.4\times 10^{16}$ atoms cm$^{-2}$, and into (b) alumina with ion energy $40$ keV and 
dose $\varphi=2.5\times 10^{16}$ atoms cm$^{-2}$. With these doses, the systems are below their respective percolation 
thresholds, presenting isolated gold nanoparticles. The arrows in (b) delimit $20$ nm that is the conducting layer thickness considered.}\label{fig2}
\end{center}
\end{figure}

The fitted values of the transport exponents 
are both consistent with the 2d value $t\simeq 1.3$, indicating that ion-implanted Au-PMMA and Au-alumina composites behave 
as 2d percolating composites for $x>x_c$. This result is consistent with the values of the thickness of the conducting layers 
of Au-PMMA and Au-alumina extracted from the TRYDIN analysis and from  cross sectional TEM images of Au-PMMA. 
For the case of Au-PMMA films we observed a conducting layer of 
thickness $h\approx 5.5-8$ nm (see Fig.~\ref{fig2}(a)) formed by Au particles of size $D\approx5-6$ nm,\cite{Teixeira2009} which indicates that 
Au-PMMA films are composed basically of one monolayer of Au particles embedded into the PMMA substrate. 
Au-implanted PMMA films are thus strictly 2d systems. On the contrary, the conducting layer of Au-alumina films is 
not strictly two-dimensional: the mean Au particle size extracted from TEM is $D\approx3.2$ nm,\cite{Salvadori2013} while 
TRYDIN analyses indicate that $h\approx 20$ nm (see Fig.~\ref{fig2}(b)), that is, about $6$ times larger than $D$. The 2d character 
of the percolation conductivity of Au-alumina films is explained by observing that the relevant length scale that governs 
the power-law behavior of Eq. (1) is the correlation length $\zeta$ (not to be confused with the tunneling decay length $\xi$), 
which measures the typical size of finite clusters of connected particles.\cite{Stauffer1994} 
When $x$  approaches $x_c$  from either above or below the percolation threshold  $x_c$, the correlation length increases 
as  $\zeta\approx a\vert x-x_c\vert^{-\nu}$, where  $a$ is of the order of the particle size $D$, and $\nu>0$  is the correlation length exponent. 
For concentrations such that $\zeta$  becomes larger than the film thickness  $h$, 
the composite behaves effectively as a 2d system,\cite{Sotta2003,Zekri2011} so that in the vicinity of $x_c$ the conductivity 
is expected to follow Eq.~\eqref{power} with $t\simeq 1.3$, as observed in our Au-alumina samples.\cite{Note2}

\begin{figure}[t]
\begin{center}
\includegraphics[scale=0.38,clip=true]{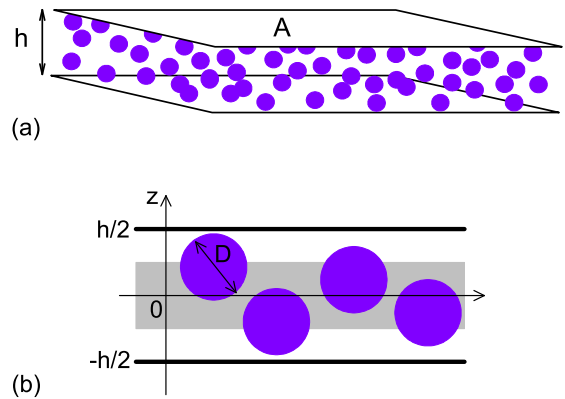}
\caption{(Color online) Model of a nanocomposite film with finite thickness of the conducting layer. (a) The metallic particles 
are taken to be spheres with identical diameter $D$ dispersed within the space delimited by two parallel hard planes of area $A$ 
and separated by a distance $gh$. The spheres cannot penetrate the confining walls. (b) Cross-sectional view illustrating the 
volume $V_h=(h-D)A$ available for the sphere centers (grey region).}\label{fig3}
\end{center}
\end{figure}

Concerning the percolation thresholds extracted from the fits, we note that the values of $x_c$  of Au-PMMA and Au-alumina are 
basically the same, although the percolation threshold is generally expected to depend on the particular combination of metal 
and insulator components constituting the films and on the details of the fabrication process.\cite{Abeles1975,Grimaldi2014} We have 
no arguments ruling out that the equivalence of the two percolation thresholds is just a mere coincidence, as very similar 
thresholds have already been observed in granular films with different components.\cite{Grimaldi2014} Interestingly, however, 
the value $x_c\simeq 0.44$  of our Au-alumina films is larger than the percolation threshold of three-dimensional Au-alumina composites 
($x_c\simeq 0.38$), which is consistent with the observation that the percolation threshold generally diminishes as the dimensionality 
increases.\cite{Sahimi2003}

For values of $x$ lower than $x_c$ clusters of coalesced particles no longer span the entire composite layer and the conductivity 
is dominated by tunneling processes, which give rise to the hump of $\sigma$ at $x<x_c$ in Fig.~\ref{fig1}. In this region the 
conductivities of Au-PMMA and of Au-alumina films are expected to follow Eq.~\eqref{tun1} with typical inter-particle distances 
$\delta(x)$ whose $x$-dependence is influenced by the effective dimensionality of the system. Although for the case of Au-PMMA films, 
which are strictly two-dimensional, $\delta(x)$ can be determined by considering dispersions of conducing particles in a plane, 
for Au-alumina films one must consider the effect of the finite width $h$ of the conducting layer, as its observed value ($h\approx 6D$)
is such that tunneling processes between Au particles at different depths within the conducting layer are also possible. 

\begin{figure}[t]
\begin{center}
\includegraphics[scale=0.33,clip=true]{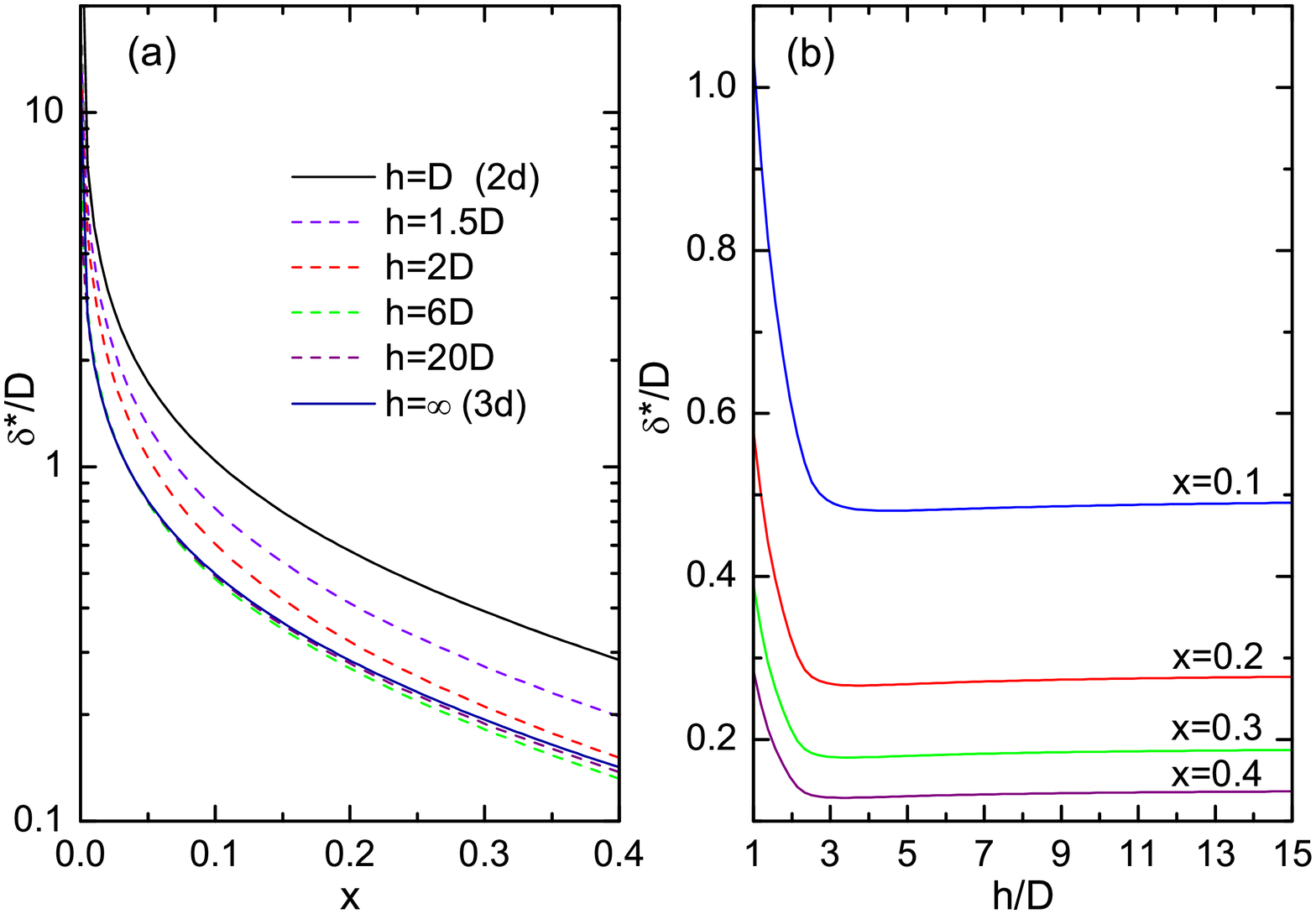}
\caption{(Color online) (a) Relevant tunneling distance $\delta^*$ between the surfaces of two spheres as a function 
of the metallic volume fraction $x$ and for different thicknesses $h$ of the film. 
$\delta^*$ is calculated from $\delta^*=(\xi/2)\ln(1/g^*)$, where $g^*$ is the numerical solution of the EMA 
equations \eqref{ema4} and \eqref{ema5} with tunneling decay length fixed at $\xi=0.05 D$. Solid lines are results 
for the limits 2d and 3d given in Eqs.~\eqref{delta1} and \eqref{delta2}. (b) $\delta^*$ as a function of $h$ for selected 
values of the volume fraction of the metallic spheres. For $\delta^*\gtrsim 3D$ the films behave practically as 3d systems.}\label{fig4}
\end{center}
\end{figure}

To tackle this problem, we consider an effective medium approximation (EMA) for the tunneling conductivity 
applied to a dispersion of $N$ spherical particles of diameter $D$ that are confined by two parallel hard walls separated by a 
distance $h\geq D$ and of macroscopic area $A$, as shown schematically in Fig.~\ref{fig3}(a). The EMA conductivity $\bar{g}$ is the solution 
of the following equation:\cite{Ambrosetti2010b,Grimaldi2014}
\begin{equation}
\label{ema1}
\frac{1}{N}\left\langle \sum_{i\neq j}\frac{g(r_{ij})}{\bar{g}+g(r_{ij})}\right\rangle=2,
\end{equation}
where $g(r_{ij})=g_0\exp[-2(r_{ij}-D)/\xi]$, with $r_{ij}\geq D$, is the tunneling conductance between two spherical particles $i$ and $j$, 
$r_{ij}=\vert\vec{r}_i-\vec{r}_j\vert$ is the distance between their centers located at $\vec{r}_i$ and $\vec{r}_j$,
and $g_0$ is a tunneling prefactor. The angular brackets in Eq.~\eqref{ema1} denote a statistical average over the 
positions of the spheres occupying the volume delimited by the two parallel planes located at 
$z=\pm h/2$, as shown in Fig.~\ref{fig3}(b). Since the spherical particles cannot penetrate the hard walls, the available volume for
the sphere centers is $V_h=(h-D)A$.
Assuming that the particles are uncorrelated, the average reduces to:
\begin{equation}
\label{ave}
\left\langle\left(\cdots\right)\right\rangle=\frac{1}{V_h^N}\int_{V_h}d\vec{r}_1\cdots\int_{V_h} d\vec{r}_N\left(\cdots\right),
\end{equation}
so that Eq.~\eqref{ema1} can be rewritten as:
\begin{equation}
\label{ema2}
\frac{N}{V_h^2}\int_{V_h}d\vec{r}_1\int_{V_h}d\vec{r}_2\frac{\theta(r-D)}{g^*\exp[2(r-D)/\xi]+1}=2,
\end{equation}
where $g^*=\bar{g}/g_0$, $\theta(x)=1$ for $x\geq 0$ and $\theta(x)=0$ for $x<0$, and $r=\vert\vec{r}_1-\vec{r}_2\vert$.
Using
\begin{equation}
\label{sumz}
\iint_{-(h-D)/2}^{(h-D)/2}\!dz_1\,dz_2=\int_{-\infty}^{+\infty}\!dz\,(h-D-\vert z\vert)\theta(h-D-\vert z\vert),
\end{equation}
where $z=z_1-z_2$, we express Eq.~\eqref{ema2} in terms of an integral over $\vec{r}=\vec{r}_1-\vec{r}_2$ which, due
to the exponential decay of the tunneling conductance, can be extended over the whole space:
\begin{equation}
\label{ema3}
\frac{h\rho}{h-D}\int\! d\vec{r}\, \theta(r-D)\frac{(h-D-\vert z\vert)\theta(h-D-\vert z\vert)}{g^*\exp[2(r-D)/\xi]+1}=2,
\end{equation}
where $\rho=N/(hA)$ is the particle number density. Finally, we pass to spherical coordinates and integrate over the angles 
to find that the EMA conductivity $g^*$ satisfies:
\begin{equation}
\label{ema4}
\frac{12 h\eta}{D^3}\int_D^{\infty}\!dr\,\frac{r}{g^*\exp[2(r-D)/\xi]+1}=2
\end{equation}
for $D\leq h\leq 2D$, and
\begin{align}
\label{ema5}
&\frac{12 h\eta}{D^3}\int_{h-D}^{\infty}\!dr\,\frac{r}{g^*\exp[2(r-D)/\xi]+1}\nonumber\\
&+\frac{24 h\eta}{(h-D)D^3}\int_D^{h-D}\!dr\,\frac{r^2}{g^*\exp[2(r-D)/\xi]+1}\nonumber\\
&-\frac{12 h\eta}{(h-D)^2D^3}\int_D^{h-D}\!dr\,\frac{r^3}{g^*\exp[2(r-D)/\xi]+1}=2
\end{align}
for $h\geq 2D$. In Eqs.~\eqref{ema4} and \eqref{ema5} we have introduced the dimensionless density $\eta=\pi D^3\rho/6$ which, 
for the case of uncorrelated (i.e., penetrable) spheres, is related to the volume fraction $x$ of the metallic phase through
$x=1-\exp(-\eta)$.\cite{Torquato1990} The 2d and 3d limits are obtained by setting, respectively, $h=D$ in Eq. (8) and
$h\gg D$ in Eq. (9). It is instructive to express $g^*$ as a tunneling conductance between the surfaces of two spheres 
separated by a characteristic distance $\delta^*$:
\begin{equation}
\label{tun2}
g^*=e^{-\frac{2\delta^*}{\xi}}.
\end{equation}   
In this way, the term $1/\{g^*\exp[2(r-D)/\xi]+1\}$ that appears in the integrands of Eqs.~\eqref{ema4} and \eqref{ema5} reduces
for $\xi/D\ll 1$ to:
\begin{align}
\label{tun3}
\frac{1}{g^*\exp[2(r-D)/\xi]+1}&=\frac{1}{\exp[2(r-D-\delta^*)/\xi]+1}\nonumber\\
&\approx\theta(\delta^*+D-r).
\end{align}
Using Eq.~\eqref{tun3} in Eq.~\eqref{ema4}, we find thus that the characteristic distance $\delta^*$ for a strictly
2d system becomes:
\begin{equation}
\label{delta1}
\frac{\delta^*}{D}=\left[\frac{1}{3\ln(1-x)^{-1}}+1\right]^{1/2}-1,
\end{equation}
while, since the first and third integrals in Eq.~\eqref{ema5} vanish for $h\gg D$, for a 3d system we obtain:
\begin{equation}
\label{delta2}
\frac{\delta^*}{D}=\left[\frac{1}{4\ln(1-x)^{-1}}+1\right]^{1/3}-1.
\end{equation}
For $x\ll 1$ the above expressions correctly reproduce the dimensional scaling $\delta^*\propto D/x^{1/d}$ expected to hold 
true in the dilute limit. Furthermore, Eqs.~\eqref{delta1} and \eqref{delta2} evidence that, for moderately small values of $x$, 
the relevant tunneling distance for 2d systems is about twice that for 3d systems, as illustrated in Fig.~\ref{fig4}(a) where 
Eqs.~\eqref{delta1} and \eqref{delta2} are shown by solid lines.

To assess how the EMA conductivity evolves from the 2d to the 3d limits, we solve numerically Eqs.~\eqref{ema4} and \eqref{ema5} 
for films thicknesses ranging from $h=D$ to $h\gg D$.  Expressing the resulting $g^*$ in terms of the distance $\delta^*$ as 
specified in Eq.~\eqref{tun2}, we find that $\delta^*$ rapidly decreases as the film thickness increases from $h=D$, 
and essentially matches the 3d limit already for $h\gtrsim 3D$, as shown in Fig.~\ref{fig4}(b) where $\delta^*$ is presented as a 
function of $h/D$ for selected $x$ values. The rapid evolution from 2d to 3d is also illustrated in Fig.~\ref{fig4}(a), where the 
numerical values of $\delta^*$ as function of $x$ and for several film thicknesses are compared with Eqs.~\eqref{delta1} and 
\eqref{delta2}. 

\begin{figure}[t]
\begin{center}
\includegraphics[scale=0.33,clip=true]{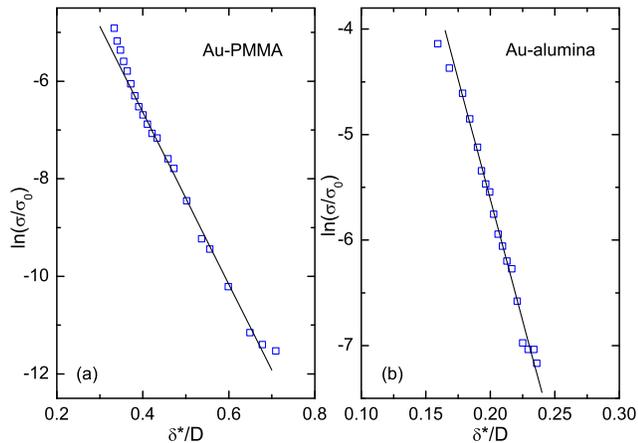}
\caption{(Color online) Natural logarithm of measured $\sigma/\sigma_0$ (open squares) as a function of $\delta^*/D$ 
for (a) Au-PMMA and (b) Au-alumina nanocomposite films. The characteristic EMA distance $\delta^*$ is assumed to be given by 
the 2d limit of Eq.~\eqref{delta1} for Au-PMMA and by the 3d limit of Eq.~\eqref{delta2} for Au-alumina. 
Solid lines are the best fits of the data with Eq.~\eqref{fit}. }\label{fig5}
\end{center}
\end{figure}

The above analysis suggests that it suffices to have film thicknesses only a few times larger than the particle size to 
induce a 3d character to the tunneling conductivity. This result is consistent with the observation that the relevant 
length scale for tunneling is given by the mean inter-particle distance, and has interesting consequences with respect 
to the nanocomposite films considered here. Indeed, if on the one hand Au-PMMA films are expected to display 2d conductivity 
in the whole range of $x$ because $h\approx D$, on the other hand the conductivity of Au-alumina nanocomposites should be 
understood as having 2d character for $x>x_c$ but 3d character for $x<x_c$. This is so because the measured thickness of 
the conducting layer of Au-alumina ($h\approx 6D$) is sufficiently larger than the EMA 2d-3d crossover value $h\approx 3D$ 
(see Fig.~\ref{fig4}) to induce 3d tunneling.

Assuming that the above considerations capture the essential physics of the problem, we interpret the measured conductivity 
data of Au-PMMA and Au-alumina films in terms of Eq.~\eqref{tun1}, where $\delta(x)$ is identified with $\delta^*$ as given 
by Eqs.~\eqref{delta1} and \eqref{delta2}, respectively. Consistently with the exponential form of Eq.~\eqref{tun1}, 
the $\ln(\sigma/\sigma_0)$ data of both systems follow approximately a linear dependence as a function of $\delta^*$, 
as shown in Fig.~\ref{fig5}. From linear fits of $\ln(\sigma/\sigma_0)$ with 
\begin{equation}
\label{fit}
\ln(\sigma/\sigma_0)=-\frac{2D}{\xi}\frac{\delta}{D}+\textrm{const.},
\end{equation}
we extract the values of $2D/\xi$ that best fit the data to find $\xi/D=0.113\pm 0.002$  for Au-PMMA and $\xi/D=0.044\pm 0.001$ 
for Au-alumina. Remarkably, the value of $\xi/D$ found for Au-alumina coincides with the value extracted from analyses of 
co-sputtered Au-alumina granular thick films,\cite{Abeles1975} which are three-dimensional systems. 
Using the estimated mean size of Au particles in our films ($D\approx 3.2$ nm) we find $\xi\approx 0.14$ nm, which compares fairly 
well with $\xi\approx 0.1$ nm, obtained from $\xi=\hbar/\sqrt{2m\Delta E}$, where $m$ is the electron mass and $\Delta E\approx 4$ eV 
is the estimated barrier height for tunneling between Au and alumina.\cite{Grimaldi2014}
It is worth noting that fitting the Au-alumina conductivity data using Eq.~\eqref{fit} with $\delta^*$ as given by the 2d limit of 
Eq.~\eqref{delta1} gives $\xi\approx 0.3$ nm, which is about twice the value found assuming 3d tunneling.
From $\xi/D\simeq 0.113$ found for Au-PMMA films and from the corresponding mean Au particle size ($D\approx 5-6$ nm) we find 
$\xi\approx 0.6-0.7$ nm, which indicates that the tunneling decay length of Au-PMMA is substantially larger than that 
of Au-alumina nanocomposites.

\begin{figure}[t]
\begin{center}
\includegraphics[scale=0.38,clip=true]{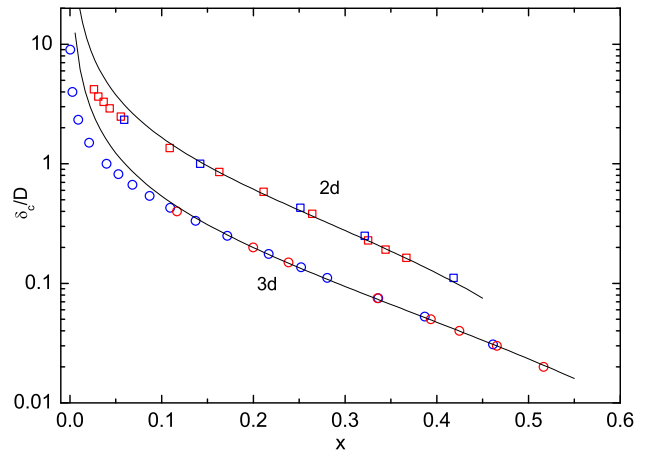}
\caption{(Color online) Monte Carlo results (open symbols) for the critical distance $\delta_c$ calculated for dispersions 
of impenetrable disks in 2d,\cite{Bug1985,Lee1990} and impenetrable spheres in 3d.\cite{Ambrosetti2010a,Miller2009} 
Solid lines are $\delta_c=2.5(r_{NN}^{2d}-D)$ for the 2d case and $\delta_c=1.65(r_{NN}^{3d}-D)$ for the 3d case, where 
$r_{NN}^{2d}$ and $r_{NN}^{3d}$ are the expressions of the corresponding mean distances between the centers of nearest 
neighboring particles given in Eqs.~\eqref{r2d} and \eqref{r3d}.}
\label{fig6}
\end{center}
\end{figure}

To assess the robustness of the results based on the EMA approach, we repeat the above analysis by identifying $\delta$ of 
Eq.~\eqref{tun1} with the critical distance $\delta_c$, as prescribed by the critical path 
approximation (CPA).\cite{Ambegaokar1971,Hunt2005,Ambrosetti2010a,Chatterjee2013} 
According to CPA, $\delta_c$  is defined as the shortest among the inter-particle distances  $\delta_{ij}=r_{ij}-D$ such that 
the set of bonds satisfying $\delta_{ij}\leq \delta_c$ forms a percolating cluster.
For dispersions of metallic particles in which the distances $\delta_{ij}$ span several 
multiples of $\xi$, CPA ensures that Eq.~\eqref{tun1} with 
$\delta=\delta_c$ gives a good estimate of the composite conductivity.\cite{Ambrosetti2010a} 

\begin{figure}[t]
\begin{center}
\includegraphics[scale=0.33,clip=true]{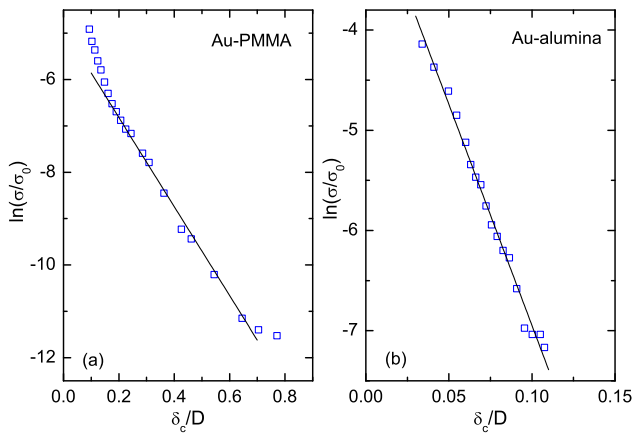}
\caption{(Color online) Natural logarithm of measured $\sigma/\sigma_0$ (open squares) as a function of $\delta_c$ for (a) 
Au-PMMA and (b) Au-alumina nanocomposite films. The critical distance $\delta_c$ is assumed to be given by the 2d limit 
for Au-PMMA and by the 3d limit for Au-alumina. Solid lines are the best fits of the data with Eq.~\eqref{fit}. }
\label{fig7}
\end{center}
\end{figure}

Monte Carlo results of $\delta_c$ for impenetrable 
spheres dispersed in a 3d volume\cite{Ambrosetti2010a,Miller2009} and impenetrable disks dispersed in a 2d 
area\cite{Bug1985,Lee1990} are shown in Fig.~\ref{fig6} by open symbols. For the 2d case, we set $x=2\phi_{2d}/3$ to convert the area 
fraction $\phi_{2d}$ covered by the disks to the volume fraction $x$ of spheres of equal diameter with centers lying on a plane. 
In analogy with the EMA results of Fig.~\ref{fig4}, the critical distance for the 2d case is systematically larger than the critical 
distance in 3d. For $x$ larger than about $0.1$ the Monte Carlo data are well reproduced by setting
$\delta_c=2.5(r_{NN}^{2d}-D)$ for the 3d case and $\delta_c=1.65(r_{NN}^{3d}-D)$ for the 3d case (solid lines in Fig.~\ref{fig6}), where 
\begin{equation}
\label{r2d}
r_{NN}^{2d}=D+D\frac{(1-3x/2)^2}{6x(2-3x/2)},
\end{equation}
and
\begin{equation}
\label{r3d}
r_{NN}^{3d}=D+D\frac{(1-x)^3}{12x(2-x)},
\end{equation}
are the mean distances between the centers of nearest neighboring spheres in 2d and 3d, respectively.\cite{Torquato1990,MacDonald1992} 
We fit the $\ln(\sigma/\sigma_0)$ data of Au-PMMA and Au-alumina with Eq.~\eqref{fit} using respectively the 2d and 3d functional 
dependences of $\delta_c$, as shown in Fig.~\ref{fig7}. From the slopes of the straight lines we extract $\xi/D=0.208\pm 0.004$ for Au-PMMA 
and $\xi/D=0.045\pm 0.001$ for Au-alumina. We immediately see that for the Au-alumina films the estimates of $\xi/D$ from CPA and 
EMA coincide within errors, while those for Au-PMMA films differ by almost a factor $2$. This discrepancy could be attributed to 
possible effects of local particle correlations, neglected within our EMA approach but fully accounted for in CPA, which are 
expected to be more prominent in 2d than in 3d.  Combining the estimates from EMA and CPA, and using the measured mean size 
of Au particles, we infer that in 2d Au-PMMA films the tunneling decay length is $\xi\approx 0.6-1.2$ nm.

\section{Conclusions}
\label{concl}

We have presented a study of the dependence of the electrical conductivity on the gold concentration in Au-implanted composite 
thin films with different insulating matrices. We have evidenced that the film thickness may influence substantially the 
behavior of the conductivity below the percolation threshold $x_c$, where tunneling between isolated gold particles dominates. 
Specifically, we have shown that an effective medium theory predicts a crossover from two-dimensional to three-dimensional 
tunneling behavior when the film thickness $h$ is larger than only about 3 times the mean Au particle size $D$. Au-implanted 
PMMA films, that have thickness$h\approx D$, are thus strictly 2d systems in the tunneling regime, while Au-implanted alumina 
films are expected to show 3d tunneling behavior, as for this system $h\approx 6D$. Interestingly, the dimensionless tunneling 
decay length $\xi/D$ extracted from the tunneling conductivity data of Au-implanted alumina films coincides with previous 
estimates of $\xi/D$ in co-sputtered Au-alumina thick film composites. We have also shown that above the percolation threshold 
the measured conductivity of both Au-PMMA and Au-alumina follows a percolation power-law behavior with 2d transport exponent, 
in accord with the theory of percolation in thin disordered films. Au-PMMA films have thus 2d character in the whole range of 
Au concentrations, while the effective dimensionality of Au-implanted alumina films increases from 2d to 3d as the system crosses 
over from the percolation regime to the tunneling regime. 
These results and interpretations could find a firmer confirmation by measuring the conductivity behavior in films 
with thicknesses ranging continuously from $h\approx D$  to  $h\gg D$. From our model we expect that the tunneling conductivity 
crosses over from two-dimensional to three-dimensional behaviors when the film thickness is about $2D$-$3D$, as inferred from the 
behavior shown in Fig.~\ref{fig4}. 

\acknowledgements
This work was supported by the Funda\c{c}\~ao de Amparo \`a Pesquisa do Estado de
S\~ao Paulo (FAPESP) and the Conselho Nacional de Desenvolvimento Cient\'ifico e
Tecnol\'ogico (CNPq), Brazil. We are grateful to the Institute of Ion Beam
Physics and Materials Research at the Forschungszentrum Dresden-Rossendorf,
Germany, for the TRIDYN-FZR computer simulation code.

\end{document}